\documentstyle[epsf,twoside,fleqn,espcrc2]{article}

\newcommand{\AmS}{{\protect\the\textfont2
  A\kern-.1667em\lower.5ex\hbox{M}\kern-.125emS}}

\def\simm#1{\mathop{\vtop{\ialign{##\crcr
        $\hfil\displaystyle{#1}\hfil$\crcr\noalign{\kern0.5pt\nointerlineskip}
        $\sim$\crcr\noalign{\kern0.5pt}}}}\limits}

\title{The CP-PACS Project}
\author{
Y.~Iwasaki\rlap,
\address{Center for Computational Physics
and Institute of Physics, University of Tsukuba, Ibaraki 305, Japan}\\
for the CP-PACS collaboration
}

\begin{document}
\renewcommand{\textfraction}{0.1}
\renewcommand{\topfraction}{0.9}
\begin{abstract}
The CP-PACS project is a five year plan, which 
formally started in April 1992 and has been completed in March 1997,
to develop a massively parallel computer
for carrying out research in computational physics with primary emphasis
on lattice QCD.
The initial version of the CP-PACS computer
with a theoretical peak speed of 307 GFLOPS with 1024 processors
was completed
in March 1996.
The final version with a
peak speed of 614 GFLOPS with 2048 processors
was completed in September 1996, and has been in full operation since 
October 1996.
We describe the architecture,
the final specification,
the hardware implementation, and
the software of the CP-PACS computer.
The CP-PACS has been used for hadron spectroscopy production runs
since July 1996.
The performance for lattice QCD applications and the LINPACK
benchmark are given.
\end{abstract}

% typeset front matter (including abstract)
\maketitle
\setcounter{footnote}{0}
\section{Introduction}
Numerical studies of lattice QCD have developed significantly
during the past decade in parallel with the development of computers.
Of particular importance in this regard has
been the construction of dedicated QCD computers (see for reviews 
Ref.\cite{review}) and
the move of commercial vendors toward parallel computers in recent years. 
In Japan the first dedicated QCD computer was developed in the QCDPAX
project~\cite{QCDPAX}.
The QCDPAX computer with a peak speed of 14GFLOPS 
is actually the 5th computer in the PAX project~\cite{PAX},
which pioneered the development of parallel computers for scientific and
engineering applications in Japan.

The CP-PACS project was conceived as a successor of the QCDPAX project in 
the early summer of 1991.
The project name CP-PACS is an acronym for Computational
Physics by a Parallel Array Computer System.
The aim of the project was
to develop a massively parallel computer
for carrying out research in computational physics with primary emphasis
on lattice QCD.

The CP-PACS project 
started in April 1992,  and after 5 years, is coming to a conclusion
in March 1997.
Therefore it is timely to overview the CP-PACS project and the CP-PACS
computer
at this workshop held in the middle of March 1997.
In this article we present an overview of the chronology and the 
organization of the CP-PACS project in Sec.2, and describe the details of
the CP-PACS computer including the architecture,
the final specification, the hardware implementation, and
the software in Sec.3.
Research areas which are covered by the CP-PACS project are
given in Sec.4.
In Sec.5 the performance of the computer
for lattice QCD applications as well as for the LINPACK benchmark are
given. 
Physics results obtained on the CP-PACS computer are presented in other
contributions~\cite{yoshie,kanaya}.
Sec.6 is devoted to conclusions.

\section{The CP-PACS Project}

The CP-PACS project~\cite{cppacs} aims at developing a massively parallel 
computer designed
to achieve high performance for numerical research of the major problems of
computational physics.
It further aims at significant
progress in the solution of these problems through the application of the 
computer upon completion of its development.

\begin{table*}[t]
\setlength{\tabcolsep}{1.0pc}
\newlength{\digitwidth} \settowidth{\digitwidth}{\rm 0}
\catcode`?=\active \def?{\kern\digitwidth}
\caption{CP-PACS Project members}
\label{tab:member}
\begin{center}
%\begin{tabular*}{\textwidth}{lllll}
\begin{tabular}{lllll}
\hline
\multicolumn{2}{c}{computer science} 
&\multicolumn{3}{c}{computational physics}\\
\hline
 hardware        & software     & particle physics 
 & astrophysics  & condensed matter\\
\hline
 K. Nakazawa${}^a$      & I. Nakata${}^e$          & Y. \
Iwasaki${}^{c}$   & S. Miyama${}^o$  & S. Miyashita${}^p$\\
 H. Nakamura${}^b$	& Y. Yamashita${}^e$	& A. Ukawa${}^k$	  & T. Nakamura${}^l$	& M. Imada${}^q$\\ 
 T. Boku${}^c$	    & Y. Oyanagi${}^f$	  & K. Kanaya${}^c$	 & M. Umemura${}^c$
& K. Nemoto${}^r$\\
 T. Hoshino${}^d$	 & T. Kawai${}^g$	    & S. Aoki${}^k$	   &
Y. Nakamoto${}^c$  &  A. Oshiyama${}^k$\\ 
 T. Shirakawa${}^d$& M. Mori${}^h$	     & T. Yoshie${}^c$ &   &  S. Gunji${}^c$\\
 K. Wada${}^e$	    & Y. Watase${}^i$	   & M. Okawa${}^l$\\
 M. Yasunaga${}^e$  & S. Ichii${}^j$     & N. Ishizuka${}^c$\\
 S. Sakai${}^c$  &        & M. Fukugita${}^m$\\
                  &                   & H. Kawai${}^n$\\
\hline
\multicolumn{5}{l}{\small ${}^a$ Department of Computer Science, University of Electro-Communications}\\[-2pt] 
\multicolumn{5}{l}{\small ${}^b$ Center for Advanced Science and Techology, University of Tokyo}\\[-2pt] 
\multicolumn{5}{l}{\small ${}^c$ Center for Computational Physics, University of Tsukuba}\\[-2pt] 
\multicolumn{5}{l}{\small ${}^d$ Institute of Engineering Mechanics, University of
Tsukuba}\\[-2pt] 
\multicolumn{5}{l}{\small ${}^e$ Institute of Information Sciences and Electronics,
University of Tsukuba}\\[-2pt] 
\multicolumn{5}{l}{\small ${}^f$ Department of Information Science, University of
Tokyo}\\[-2pt] 
\multicolumn{5}{l}{\small ${}^g$ Department of Physics, Keio University}\\[-2pt] 
\multicolumn{5}{l}{\small ${}^h$ Department of Engineering, University of
Tokyo}\\ [-2pt]
\multicolumn{5}{l}{\small ${}^i$ Data Handling Division, KEK}\\[-2pt] 
\multicolumn{5}{l}{\small ${}^j$ Computer Center, University of Tokyo}\\[-2pt] 
\multicolumn{5}{l}{\small ${}^k$ Institute of Physics, University of Tsukuba}\\[-2pt] 
\multicolumn{5}{l}{\small ${}^l$ Numerical Theory Division, KEK}\\ [-2pt]
\multicolumn{5}{l}{\small ${}^m$ Yukawa Institute for Theoretical Physics,
Kyoto University}\\[-2pt] 
\multicolumn{5}{l}{\small ${}^n$ Theory Division, KEK}\\ [-2pt]
\multicolumn{5}{l}{\small ${}^o$ National Astronomial Observatory}\\[-2pt] 
\multicolumn{5}{l}{\small ${}^p$ Department of Physics, Osaka University}\\[-2pt] 
\multicolumn{5}{l}{\small ${}^q$ Institute of Solid State Physics, University of Tokyo}\\[-2pt] 
\multicolumn{5}{l}{\small ${}^r$ Department of Physics, Hokkaido University}\\[-2pt]  
\end{tabular}
\vspace{-0.3 cm}
\end{center}
\end{table*}

The planning of the project was started in the summer of 1991. The proposal,
made to the Ministry of Education, Science and Culture,
was approved in the spring of 1991 as one of projects of the Ministry's
``Program for New Development of Academic Research''. The project
formally started in April of 1992,
and has received about 2.2 billion yen spread 
over the five year period ending in March 1997.
The funding comes from a special allocation of the Grant-in-Aid of the
Ministry of Education, Science and Culture supporting innovative fundamental
research.

The Center for Computational Physics was founded in April 1992 at University
of Tsukuba to carry out the project, as well as to promote
research in computational physics and parallel computer science. The Center
is an inter-university facility open to researchers in academic institutions
in Japan.
% even after the project formally ends.

The number of the project members, which was 22 when the project started,
has increased to 33, of which 15 are computer
scientists and 18 are physicists, as listed in Table~\ref{tab:member}.
The projected was headed by Y. Iwasaki. The development of the CP-PACS
computer was led by K. Nakazawa.

A unique feature of the project,
as is clear from Table~\ref{tab:member},
is its emphasis on cross-disciplinary
research involving both physicists and computer scientists. This is a 
tradition carried over from the QCDPAX project~\cite{QCDPAX},
which is the predecessor and  stepping stone for the CP-PACS project. 
A close collaboration of researchers from
the two disciplines has been both important and fruitful in reaching a
design for the CP-PACS computer which best balances the computational needs 
of physics applications with the latest of computer technologies.

Development of a massively parallel computer requires advanced semiconductor 
technology.  We discussed the aim of the project with a number of 
manufacturers and invited proposals
in the period of 1991--1992.  
We selected Hitachi Ltd. as the industrial parter through a formal
bidding process in the early summer of 
1992, and we
have worked in a close collaboration for the hardware and software
development of the CP-PACS computer.
The fundamental design of the computer was laid down
in 1992, its details worked out in 1993, the logical design and
the physical packaging design completed in 1994, and
chip fabrication and assembling of parts started in early 1995.
The first stage of the CP-PACS computer consisting of 1024 processing
units with a peak speed of 307 GFLOPS was completed in March 1996. An upgrade
to a 2048 system with a peak speed of 614GFLOPS
was completed at the end of September 1996

\section{CP-PACS Computer}

\subsection{Architecture}

The CP-PACS computer is an 
MIMD (Multiple Instruction-streams Multiple Data-streams)
parallel computer with a theoretical peak speed of 614GFLOPS and a
distributed memory of 128 Gbyte. The system consists of 2048
processing units (PU's) for parallel floating point processing and 128
I/O units (IOU's) for distributed input/output processing. These units
are connected in an 8$\times$17$\times$16 
three-dimensional array by a three-dimensional crossbar network. 
The specification of the CP-PACS computer is summarized in
Table~\ref{tab:specification}.

\begin{table}[tb]
\setlength{\tabcolsep}{0.1pc}
\caption{Specification of the CP-PACS computer}
\label{tab:specification}
%\vspace{-3mm}
%\begin{center}
\begin{tabular}{ll}
\hline
peak speed &614Gflops(64 bit data)\\
main memory & 128GB\\
parallel architecture&MIMD with \\
                     &distributed memory\\
number of nodes&2048\\
node processor & HP PA-RISC1.1+PVP-SW\\
\hspace*{2mm}\#FP registers&128\\
\hspace*{2mm}clock cycle & 150MHz\\
\hspace*{2mm}1st level cache&16KB(I)+16KB(D)\\
\hspace*{2mm}2nd level cache&512KB(I)+512KB(D)\\
network&3-d crossbar\\
\hspace*{2mm}node array &$8\times 17\times 16^*$\\
\hspace*{2mm}through-put&$$300MB/sec\\
\hspace*{2mm}latency&$2.5 \sim 3.1\,\mu$sec\\
distributed disks&3.5" RAID-5 disk\\
\hspace*{2mm}total capacity&1059GB\\
software&\\
\hspace*{2mm}OS&UNIX, micro kernel\\
\hspace*{2mm}language&FORTRAN, C, assembler\\
Size &7.0m(width) $\times$ 4.2m(depth)\\
 & $\times$ 2.0m(hight)\\
Power dissipation&275 KW maximum\\
\hline
\multicolumn{2}{r}{${}^*$including nodes for disk I/O}
\end{tabular}
%\end{center}
%\vspace{-10mm}
\end{table}

The basic strategy we have adopted for the design is the usage of a fast RISC
micro-processor for high arithmetic performance at each node and a 
linking of nodes with a flexible network so as to be able to handle a wide 
variety of problems in computational physics.  The
unique features of the CP-PACS computer reflecting these goals are 
represented by the special node processor architecture 
called {\it pseudo vector processor based on slide-windowed
registers} ({\it PVP-SW})~\cite{slidewindow} and the choice of a 
three-dimensional {\it Hyper Crossbar} network.  
A well-balanced performance of CPU, network and I/O devices
supports the high capability of CP-PACS for massively parallel processing.

\subsection{Node Processor}

%\subsubsection{PVP-SW\protect~\cite{slidewindow}}
%\label{sec:pvpsw}
Each PU of the CP-PACS has a custom-made superscalar RISC processor with
an architecture based on PA-RISC 1.1. In large scale computations in 
scientific and engineering applications on a RISC processor, the performance 
degradation occurring when the data size exceeds the cache memory capacity is 
a serious problem.
The PVP-SW is our solution
to this problem, while maintaining upward compatibility with
the PA-RISC architecture.

\begin{figure}[t]
\begin{center}
\leavevmode
\epsfxsize=6.0cm
\epsfbox{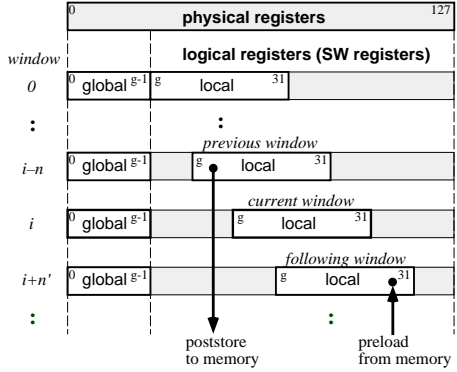}
\end{center}
\vspace{-0.2 cm}
\caption{Structure of slide-windowed registers.}
\label{fig:slidewindow}
\vspace{-0.2 cm}
\end{figure}

A schematic illustration of
the PVP-SW architecture is given in Fig.~\ref{fig:slidewindow}.
The Slide Window
mechanism allows the use of a large number of physical registers,
which is 128 in the case of CP-PACS,
through a continuously sliding logical register window of 32 registers along 
the physical registers. The Preload and Poststore
instructions can be issued without waiting for the completion
of memory access. These features
enable a pipelined access to main memory 
which is made with multiple interleaved banks, and thus a long
latency for memory access can be tolerated.  
An efficient vector processing without degradation for a very large length
of vector-loop is realized in spite of the superscalar
architecture of the CP-PACS processor.

\subsection{Network}

\begin{figure}[t]
\begin{center}
\leavevmode
\epsfxsize=6.0cm
\epsfbox{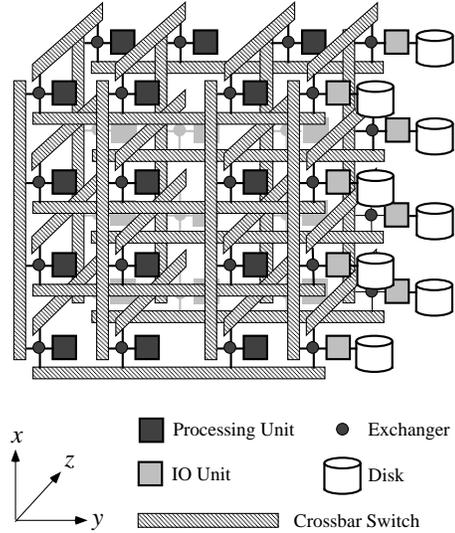}
\vspace{-0.2 cm}
\caption{Schematic diagram of CP-PACS.}  
\label{fig:cppacs}
\vspace{-0.2 cm}
\end{center}
\end{figure}

The 2048 processors are arranged in a three-dimensional $8\times 16\times 16$
array.  
The Hyper Crossbar network is made of crossbar
switches in the $x, y$ and $z$ directions, connected together by an
Exchanger at each of the three-dimensional crossing points of the
crossbar array,
as illustrated by a schematic diagram shown in
Fig.~\ref{fig:cppacs}.  
Each exchanger is connected to a PU or IOU.  Thus
any pattern of data transfer can be performed with the use of at most
three crossbar switches.
Since the network has a huge switching capacity
due to the large number of crossbar switches, the sustained data transfer
throughput in general applications is very high.

Data transfer on the network is made through Remote DMA (Remote
Direct Memory Access), in which processors exchange data
directly between their respective user
memories
with a minimum of intervention from
the operating system. This leads to a significant reduction in the startup
latency, and a high throughput.

Inter-node communication is made by message passing.  Transfer
of data within the network proceeds via wormhole routing through the 
exchangers. The direction of
routing is fixed to $x\to y\to z$ to avoid deadlocks.  
The bandwidth of each crossbar is 300Mbyte/sec. The latency, 
namely the initial overhead due to hardware and software combined,
for sending and receiving data
is 2.45, 2.83 and 3.09 $\mu$sec, respectively,
for the cases of the data transfer through one crossbar switch in the
$x$ direction, two switches in the $x$ and $y$ directions, and 
three switches in the $x, y$ and $z$ directions.

The network allows a hardware bisection of PU arrays in each of
the $x, y$ and $z$ directions.
Hence the full system can be
divided up to 8 independent subsystems.

\subsection{Distributed Disks}

The distributed disk system of CP-PACS is connected to 128 IOU's on
the $8 \times 16$ plane at the end of the $y$ direction of the Hyper Crossbar
network by a SCSI-II bus. RAID-5 disks are used for fault tolerance. The IOU's
handle parallel file I/O requests issued by the PU's
in an efficient and distributed way
using Remote DMA through the Hyper Crossbar network.

\begin{figure}[t]
\begin{center}
\leavevmode
\epsfxsize=4.5cm
  \epsfbox{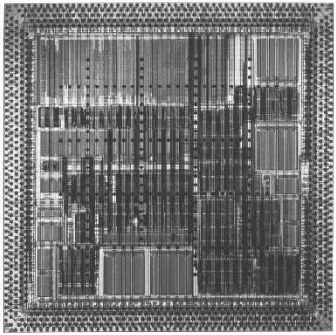}
\end{center}
\vspace{-0.2cm}
\caption{Floor-plan of the CPU chip}
\label{floor-plan}
\vspace{-0.2cm}
\end{figure}

\subsection{Connection to the Front End}
The HIPPI connection to the front host is attached to one of the IOU's. A
special FTP protocol has been developed for a high speed file transfer
between the distributed disk system of CP-PACS and the disk storage of
the front host. The peak throughput is 100 Mbytes/sec and the effective
throughput is about 65 Mbytes/sec in the case when the data with a size of
512 Mbyte are sent from CP-PACS to the front host or
from the host to CP-PACS.

\subsection{Hardware Implementation}
The CPU chip is fabricated using 0.3 micron CMOS semiconductor
technology, with the size being 15.7mm $\times$ 15.7mm. 
Fig.~\ref{floor-plan} shows the floor plan of the chip.
The PVP-SW feature is implemented with 128 floating point
registers occupying the top left block together with floating point
execution units. 

\begin{figure}[t]
\begin{center}
\leavevmode
\epsfxsize=5.5cm
  \epsfbox{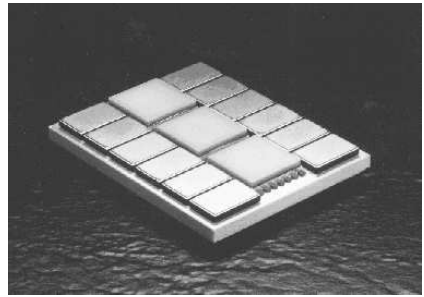}
\end{center}
\vspace{-0.2cm}
\caption{Ceramic multichip module of CPU}
\label{module}
\vspace{-0.2cm}
\end{figure}

The CPU, the storage controller (SC) and the network 
interface adapter (NIA), are mounted in-line on a ceramic multi-chip module 
of size
5.7cm $\times$ 7.2cm, which is shown in Fig.~\ref{module}:
%(three larger squares at the center).
the left one is the CPU, the central one is the SC and the right one is the 
NIA. 
The SC and NIA chips are fabricated using 0.5 micron
CMOS gate-array technology.  The twelve pieces surrounding them are the
second-level cache memory chips.

\begin{figure}[t]
\begin{center}
\leavevmode
\epsfxsize=5.5cm
  \epsfbox{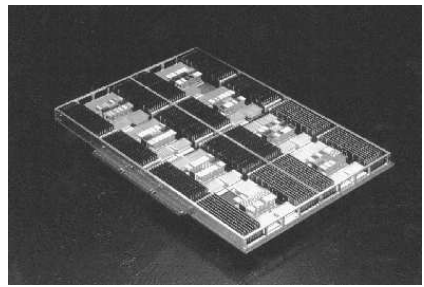}
\end{center}
\vspace{-0.2cm}
\caption{One board consists of eight CPU units}
\label{board}
\vspace{-0.2cm}
\end{figure}

Eight PU modules together with their DRAM memory are mounted on a board
of size 45.6cm $\times$ 62.5cm as shown in Fig.~\ref{board}.
The central piece of each of the eight sections
is the PU module, now with fins for air-cooling. The
other white pieces are main memory address/data control units. The black
pieces are DIM modules of 4 Mbit DRAM, 64 MByte for each PU. Each board has
two more chips for the crossbar switches in the $x$ direction, and one chip for
the clock distributer.

\begin{figure}[th]
\begin{center}
\leavevmode
\epsfxsize=5.5cm
  \epsfbox{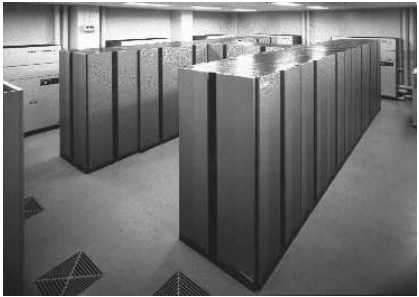}
\end{center}
\vspace{-0.2cm}
\caption{Outlook of the CP-PACS computer}
\label{outlook}
%\vspace{-0.2cm}
\end{figure}

Sixteen PU boards and one IOU board are placed vertically on a back plane,
and two back planes, one on
top of the other, are housed in a cabinet.
A Crossbar switch in the $y$ direction is mounted on the backplane
in the cabinet. Crossbar switches in the $z$ direction are
mounted on separate boards, which are housed in separate cabinets.
A picture of the CP-PACS computer is shown in Fig.~\ref{outlook}.

\begin{figure}[th]
\begin{center}
\leavevmode
\epsfxsize=6.0cm
  \epsfbox{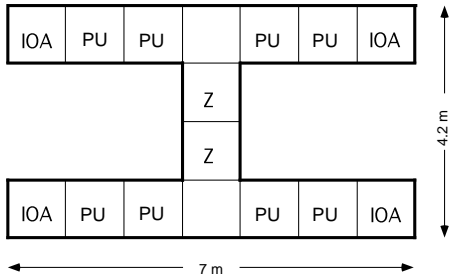}
\end{center}
\vspace{-0.2cm}
\caption{Floor-plan of the CP-PACS computer}
\label{cppacs-floor}
\vspace{-0.2cm}
\end{figure}

A schematic floor plan of the cabinets are shown in Fig.~\ref{cppacs-floor}.
The squares
labeled ``PU'' represent cabinets housing the PU and IOU's, and those labeled
``Z'' are for the crossbar switches in the $z$ direction.  The cabinets labeled
``IOA'' contain I/O adapters. The
RAID-5 distributed disk system, placed a few meters from the CP-PACS,
is connected to the IOU's by a SCSI-II bus through adapters in the IOA 
cabinets.
The system is cooled by air drawn in from beneath the cabinets.

\subsection{Software of the CP-PACS}

\subsubsection{Operating System}

The CP-PACS computer runs under the UNIX OSF/1 operating system. Each
node processor, however, carries only a kernel based on Mach 3.0
in order to save memory for user application programs and to avoid
performance degradation.  The kernel handles memory control, inter-node
communication, process scheduling, interrupt handling and I/O.  The full
UNIX interface and file server functions are implemented on the IOU's.
One of the IOU, named the SIOU, controls the whole system through the
network.

The operating system has several new functions added for parallel
processing; software partitioning of the processor array so that independent
programs may be run on different partitions, and the  generation of processes
over a user-specified number of nodes to execute a parallel program.

The file system is logically structured to form a single tree for the
entire CP-PACS computer. The file sets required to execute a single job
can be distributed over the disks connected to the parallel IOU's so as to
reduce I/O overheads. The logical and physical mapping of the file
system is automatically controlled by the operating system.

\begin{figure}[th]
\begin{center}
\leavevmode
\epsfxsize=6.5cm
\epsfbox{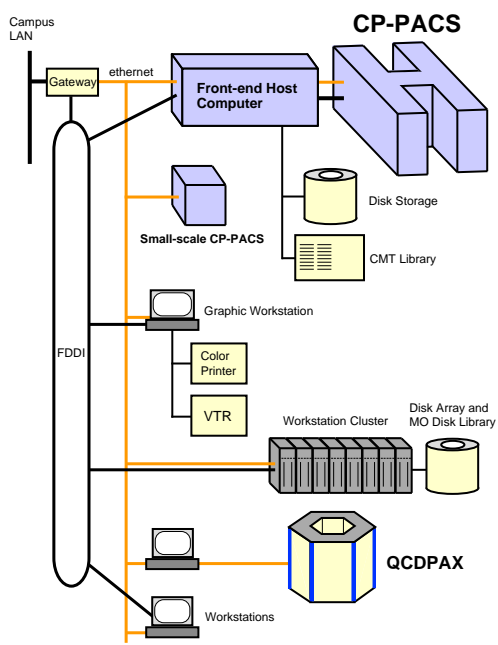}
\end{center}
\vspace{-0.2cm}
\caption{The computing system at the Center for Computational Physics.}   
\label{fig:center}
\vspace{-0.2cm}
\end{figure}

\subsubsection{Programming Environment}

FORTRAN90, C, C++ and assembly language are available for
programming on the CP-PACS computer.  Assembler code can be
included as a subroutine in a FORTRAN or C code in order to maximize the
performance.
Remote DMA data transfer through the
Hyper Crossbar network is made by calling special library routines for
communications. 
FORTRAN90 and C compilers generate assembler codes which incorporate
the PVP-SW enhancement, using
the technique of modulo scheduling and register coloring.

The Real-Time Performance Monitor allows an on-line check of the performance
of the CP-PACS in applications. Various data, including the flops
of each CPU and the busy rate of the network can be collected at regular
intervals, and can be graphically displayed on terminals.

\subsection{Front End and Mass Storage}
The computing system at the 
Center for Computational Physics
is shown in Fig.~\ref{fig:center}.
The CP-PACS computer is connected by a HIPPI channel and Ethernet to the
front host, which in turn is connected to the disk storage (350 GByte) and
a tape archive (780 GByte). The front host is a vector computer with a peak
speed of 256 MFLOPS and 1 GByte of main memory. 
Job requests for the CP-PACS are submitted through the front host using NQS.
Data I/O between the disk storage and the distributed disk system of 
the CP-PACS
is made through the HIPPI channel. Output data files are sent back to the
disk storage at the termination of each job request. 

The front host has a disk storage of 350 GByte connected by multiple channels
to achieve a high data transfer throughput. 
The front host is also connected to a magnetic cartridge tape library which
holds 980  cartridges, each with a capacity of 800 MByte.

\begin{table*}[tbh]
\caption{Performance for lattice QCD programs}
\label{tab:QCDperformance}
\begin{center}
\begin{tabular}{|c|c|c|c|c|}
\hline
program &\multicolumn{3}{c|}{ MFLOPS/PU} & coding \\
	&\multicolumn{3}{c|} {(peak 300 MFLOPS)}& \\
\hline
	&	calculation & 77\% & 191&  assembler \\
	& communication & 23\% & -&  +\\
red/black MR solver 	& sustained& 100\%&148&Fortran\\
\cline{2-5}
for Wilson quark matrix	&	calculation & 84\% & 99&    \\
	& communication & 16\% & -& Fortran\\
	& sustained& 100\%&84& \\
\hline
conjugate gradient solver&calculation & 90\% & 139& 	 \\
for Kogut-Susskind quark & communication & 10\% & -&Fortran\\
matrix& sustained& 100\%&125&  \\
\hline
Heat bath Monte Carlo&calculation & 96\% & 100&  \\
program & communication & 4\% & -&Fortran\\
for SU(3) gauge theory & sustained& 100\%&95&  \\
\hline
Over-relaxation&calculation & 91\% & 156&  \\
program & communication & 9\% & -&Fortran\\
for SU(3) gauge theory & sustained& 100\%&142&  \\
\hline
Hybrid Monte Carlo&calculation & 74\% & 151&  \\
program & communication & 26\% & -&Fortran\\
for full QCD & sustained& 100\%&112&  \\
\hline
\end{tabular}
\end{center}
\end{table*}

The Center computing facility includes a workstation cluster connected
by a high speed switch. One of the workstations functions as a file server
accessing a RAID-5 disk system with a total capacity of 89 GByte.

The QCDPAX, which was developed at University of Tsukuba by the 
QCDPAX project (1987-1990), is also a part of the system.
It has been in continuous operation since completion
for the numerical simulation of lattice QCD. 

The computing facilities of the Center are connected by a LAN consisting
of an FDDI loop and Ethernet, which in turn is connected to the University
of Tsukuba campus network.

\section{Research Areas in Computational Physics}
In computational physics the project aims to use the CP-PACS computer
for carrying out research in the following three areas: 
particle physics, condensed matter physics and astrophysics.

\begin{table*}[t]
\caption{Performance of LINPACK benchmark}
\label{tab:LINPACK}
\begin{center}
%\begin{tabular*}{10.75cm}{|r|r|r|r|r|l|}
\begin{tabular}{|r|r|r|r|r|l|}
\hline
N0. of & Rmax& Nmax& N1/2& Rpeak& Rmax\\
Procs. &(GFLOPS) &(order) &(order)&(GFLOPS) &/Rpeak(\%) \\
\hline
      1&          0.1969&       2340&         360&          0.3&         65.6\\
\hline
     2&          0.3873&       3240&         600&          0.6&         64.5\\
\hline
     4&          0.7704&       4680&         960&          1.2&         64.2\\
\hline
     8&          1.527&        6480&        1440&          2.4&         63.6\\
\hline
    16&          3.022&        9360&        2160&          4.8&         62.95\\
\hline
    32&          6.022&       12960&        3360&          9.6&         62.7\\
\hline
     64&         12.0&         18720&        4800&         19.2&         62.5\\
\hline
    128&         23.9&         25920&        6720&         38.4&         62.2\\
\hline
    256&         46.81&        37440&        9600&         76.8&         61.0\\
\hline
    512&         93.99&        51840&       15360&        153.6&         61.2\\
\hline
   1024&        186.5&         74880&       21120&        307.2&         60.7\\
\hline
   2048&        368.2&        103680&       30720&        614.4&         59.9\\
\hline
\end{tabular}
\end{center}
\end{table*}

A major goal of the project is to significantly advance the numerical study
of lattice QCD in particle physics. 
Large-scale numerical simulations will
be pursued with the CP-PACS computer in order to verify the theory and
to extract new physical predictions. 
Since the CP-PACS computer with 1024 nodes was completed, hadron spectroscopy
calculations in the quenched approximation as well as in full QCD have
been intensively performed and physics results are reported at this
workshop~\cite{yoshie,kanaya}.

Important problems in
condensed matter physics such as
strongly interacting electron systems, high-temperature
super-conductivity, first-principles calculations in material properties
and those in astrophysics such as the formation of galaxies 
and stellar/planetary 
systems, and the gravitational collapse
will be also pursued with the CP-PACS computer.
Preparations for astrophysics and condensed matter physics
applications have been started, and will gradually expand in time.
\section{Performance}
\label{qcdbenchmark}

Our codes for large-scale simulations in lattice QCD
have been written with Fortran 90 and
libraries for data communication.
The Fortran compiler which has been newly developed
to incorporate the PVP-SW feature 
produces efficient object codes,
achieving typically 90 -- 150 MFLOPS per node, depending on the structure of 
do-loops.
We have further developed a hand-optimized assembler code
for the core part of the red/black solver of the Wilson quark matrix.
In this case the performance reaches 191 MFLOPS per node which is about
64 \% of the peak speed (See  Table~\ref{tab:QCDperformance}). 
Even when the overhead due to data communication
is included,
the sustained speed in this case is 148 MFLOPS, which is about a half
of the peak speed.
The performance for typical application programs in lattice QCD is shown
in Table~\ref{tab:QCDperformance}.

We have also measured the performance for the LINPACK benchmark.
The results are summarized in Table~\ref{tab:LINPACK}.
The sustained speed for the case of 2048 PU's is 368.2 GFLOPS, 
which is 59.9\% of the theoretical peak speed.

\section{Conclusions}
We have been able to develop a massively parallel computer
of a peak speed of 614 GFLOPS through a very effective collaboration
of computer scientists, physicists and a vendor. 
Throughout the development phase
we held
a joint meeting at least once a month, discussing every aspects of the
CP-PACS computer from the architectural design
to the details of the hardware implementation.

The CP-PACS computer achieves
high performance of 40 - 50 \% of the peak speed
for lattice QCD application programs.
The machine is very stable
and we are obtaining interesting results on hadron spectrum
in the quenched QCD as well as in full QCD. 
Preparations for
astrophysics and condensed matter physics applications 
have also started.

\section*{ACKNOWLEDGEMENTS}

I would like to thank the members of the CP-PACS project,
in particular, K. Nakazawa and A. Ukawa
for valuable discussions.
I also would like to thank Hitachi Ltd. for a close collaboration
on the development of the hardware as well as the software
of the CP-PACS computer.
This work is supported in part by
the Grand-in-Aid of the  Ministry of Education, Science and 
Culture (No.~08NP0101).

\end{document}